\documentclass[11pt]{article}
\usepackage[T1]{fontenc}
\usepackage[utf8]{inputenc}
\usepackage{microtype}
\usepackage{amsmath,amssymb,bm}
\usepackage{graphicx}
\usepackage[labelfont=bf,labelsep=space]{caption}

\usepackage{geometry}
\usepackage{hyperref}
\usepackage{cite}
\hypersetup{hidelinks}

\global\long\def\tv{\hat{\boldsymbol{t}}}
\global\long\def\ten{\mathcal{T}}
\def\Nrzero{N_2}
\def\NRone{N_1}
\def\Rone{r_1}
\def\rzero{r_2}
\def\Nzero{N_2}
\def\LL{L_{0}}

\newcommand{\dd}{\mathrm{d}}
\newcommand{\order}[1]{\mathcal{O}\!\left(#1\right)}

\title{A Simple Pendulum in Schwarzschild Spacetime}
\author{Andrzej Czarnecki and Andrew Czezowski\\
Department of Physics, University of Alberta, Edmonton, Alberta, Canada}
\date{}

\begin{document}
\maketitle

\begin{abstract}
We determine the small-oscillation period of a simple pendulum in
  Schwarzschild spacetime. After
  transients decay, the displaced rope coincides with a geodesic of
  the induced spatial metric. This determines the radial lift of the
  bob to quadratic order in the angular amplitude and leads to the
  period measured by an observer at the bob's equilibrium position, $r=r_2$, with the
  fulcrum at $r_1$,
  \[
  T_{(2)}=4\pi\frac{r_2^2}{c r_s}\sqrt{N_2(N_1-N_2)},
  \]
  where $N_i=N(r_i)$ and $N(r)=\sqrt{1-r_s/r}$ is the
  Schwarzschild lapse. The result is therefore expressed in terms of
  the same function that determines clock rates and
  gravitational redshifts. In the Newtonian limit, we reproduce the
  classical result $T=2\pi\sqrt{\LL/g}$. We interpret the period close to the
  horizon in terms of an
  effective pendulum length. We compare our treatment with the
  coordinate-length constraint adopted in an earlier study and argue
  that the two problems are inequivalent away from the
  classical, weak-field limit.
\end{abstract}


\section{Introduction}

The simple pendulum is one of the most basic systems in mechanics. One might expect its small-oscillation period to be
known and readily available even in Schwarzschild spacetime. Yet the
answer is not entirely obvious. The purpose of this paper is to
provide that result: we derive the small-oscillation period, Eq.~\eqref{eq:T2}, for a
pendulum whose rope has fixed proper length $L_0$.

The problem is elementary only in appearance. In curved spacetime the
dynamics depends on how the mechanical constraint is formulated. Coordinate separations and invariant lengths
need not coincide, so the meaning of a fixed length
constraint must be specified with care. This is especially relevant in
strong fields, where the distinction between coordinate and
proper distance is significant.

Our setup is simple: the rope is taut, very light, and inextensible;
the bob is displaced slightly and held until the rope settles; the bob is then released and the period is
measured by a static observer at the bob's equilibrium position
(``observer at
the bob'' for
simplicity). 
Sect.~\ref{sec:shape} shows that  the displaced rope follows a geodesic of the
spatial metric. This leads to the period in
Eq.~\eqref{eq:T2}. 

It is useful to compare our results with Ref.~\cite{Martin-Delgado:2022nmx}
which considers a pendulum in Schwarzschild 
spacetime in the context of mechanical clocks in general 
relativity. It uses a coordinate-length 
prescription, whereas we use fixed proper length. The two models 
therefore describe different physical pendula, and comparing them 
makes clear that this seemingly simple problem is not trivial.

The notion of a static observer is not trivially generalizable to
spacetimes more complicated than Schwarzschild. For this reason, an
extension of the present study to Kerr, for example, would require
more care. For further discussion see
\cite[Sect.~3.3.3]{Gourgoulhon:2007ue}. Here we only stress that the
static property of the Schwarzschild geometry is essential in our
work.  (The spatial metric $h_{ij}$ used below is the metric induced on
hypersurfaces orthogonal to the static Killing field; this
hypersurface-orthogonal, twist-free structure is what fails in a
rotating spacetime such as Kerr.)

A further point of interest is that ropes and strings near black holes
are subtle objects. A truly massless and perfectly inextensible rope
is an idealization, and once the rope's own stress-energy is taken
into account, close-to-horizon configurations are subject to 
restrictions~\cite{Brown:2012un}. Our purpose is more modest: we use the
rope and bob as test objects.  Background material on Schwarzschild
geometry and local measurements can be found, for example, in
Refs.~\cite{PoissonWill2014,FrolovZelnikov2011}; for a reminder that
coordinates in Schwarzschild spacetime are not observables, see
Ref.~\cite{Fromholz2014}.

A static string attached to a particle in Schwarzschild spacetime was analyzed
by LaHaye and Poisson \cite{LaHaye:2021sbx}.  They considered both
massless and massive strings and addressed the
gravitational backreaction, whereas in the present work
we idealize the rope as very light and ignore its self-gravity; our
focus is instead the dynamics implied by a proper-length
constraint on the bob.

In Sect.~\ref{sec:geometry} we describe the configuration of the
pendulum and derive a relation between the coordinate rise of the bob
$\delta r$ and its angular displacement, which follows from the
fixed-length condition. In Sect.~\ref{sec:dynamics} we use that
constraint to reduce the Lagrangian of the bob and derive 
our main result, the period of small oscillations.
In Sect.~\ref{sec:limits} we examine Newtonian and close-to-horizon
limits. Sect.~\ref{sec:comparison} compares the analysis in
Ref.~\cite{Martin-Delgado:2022nmx} with ours. Finally, in
Sect.~\ref{sec:rigidity} we return to our key assumption of a
fixed-length pendulum whose shape always follows a geodesic of the
spatial metric; we demonstrate the existence of a range of
lengths in which this concept is consistent: a signal has enough time to travel
many times along the pendulum during one period. We conclude in Sect.~\ref{sec:conclusions}.

\section{Geometry of the displaced rope}
\label{sec:geometry}
This Section, divided into three subsections, starts with the notation and showing that the 
pendulum follows a geodesic of the spatial metric. In \ref{sec:ang} we 
relate the angular displacement of the bob to a constant $J$
characterizing the geodesic.  In \ref{sec:lift} we eliminate $J$ to 
express the radial lift of the bob in terms of its angular 
displacement.

\subsection{Shape of the rope when the bob is displaced}
\label{sec:shape}
\begin{figure}[htb]
\centering
\includegraphics[width=0.2\textwidth]{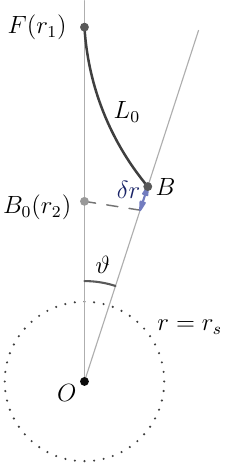}\par
\caption{Geometry of the pendulum displaced from $\theta =0$ to a
  small endpoint angle, written here as the amplitude
  $\vartheta$. $\delta r$ is the resulting change of the radial
  coordinate of the bob, $\delta r \sim \vartheta^2$. $F$ is the
  fulcrum at $r=r_1$, $B_0$ is the equilibrium position of the bob at
  $r_2$, and $B$ is the displaced bob. The dotted circle is the event
  horizon at $r=r_s$.}
\label{fig:fig1}
\end{figure}
We denote by $r$ the usual areal (also known as circumferential) radial Schwarzschild
coordinate.  We consider a rope whose upper end is fixed
at $r=r_1 $ and a bob of mass $m$ attached at its lower end,
initially at $r=r_2$, with
$r_1 >r_2>r_s$,
as shown in Fig.~\ref{fig:fig1}; $r_s =2GM/c^2$ is the Schwarzschild radius. We restrict
attention to a meridional plane $\dd\phi=0$, for which the
Schwarzschild metric is 
\begin{equation}
\dd s^{2}=-N(r)^{2} c^2 \dd t^{2}+h_{ij}\dd x^{i}\dd x^{j},
\quad
h_{ij}\dd x^{i}\dd x^{j}
=
\frac{\dd r^{2}}{N(r)^{2}}+r^{2}\dd\theta^{2},
\quad 
N(r)\equiv\sqrt{1-\frac{r_s }{r}},
\label{eq:spatialmetric}
\end{equation}
on the two-dimensional static slice containing the rope. Throughout, we write
$N_i\equiv N(r_i)$.

In the remainder of this section we set $c=1$.
 A static observer has four-velocity
\begin{equation}
u^{\mu}=\frac{1}{N}\delta_{t}^{\mu}.
\end{equation}
Let $\tv^{i}$ denote the unit tangent to the rope with respect to the spatial
metric $h_{ij}$, so that
\begin{equation}
h_{ij}\tv^{i}\tv^{j}=1.
\end{equation}
Since the rope carries only longitudinal tension, its spatial stress tensor is
\begin{equation}
S^{ij}=-\ten\,\tv^{i}\tv^{j},
\end{equation}
where $\ten$ is the tension per cross-sectional area. The stress-energy tensor
of the rope is therefore
\begin{equation}
T^{\mu\nu}=\rho u^{\mu}u^{\nu}+S^{\mu\nu},
\qquad
S^{t\nu}=0 .
\end{equation}
The following calculation starts with the stress-energy conservation
in four dimensions, uses the definition of the covariant derivative
and properties of Christoffel symbols \cite[Problem
7.7(e)]{LightmanProblemGR}, and finally leads to a conservation law in
three spatial dimensions. Consider
a spatial component of the stress-energy conservation,
\begin{align}
0 & =\nabla_{\nu}T^{i\nu}\\
 & =\partial_{\nu}T^{i\nu}+\Gamma_{\nu\beta}^{i}T^{\beta\nu}+\Gamma_{\nu\beta}^{\nu}T^{i\beta}\\
 & =\partial_{\nu}T^{i\nu}+\Gamma_{\nu\beta}^{i}T^{\beta\nu}+\frac{1}{\sqrt{-g}}T^{i\nu}\partial_{\nu}\sqrt{-g}\\
 & =\Gamma_{\nu\beta}^{i}T^{\beta\nu}+\frac{1}{\sqrt{-g}}\partial_{\nu}\left(\sqrt{-g}T^{i\nu}\right)\\
 & =\Gamma_{jk}^{i}S^{kj}+\Gamma_{tt}^{i}T^{tt}+\frac{1}{\sqrt{-g}}\partial_{j}\left(\sqrt{-g}S^{ij}\right).
\end{align}
In this section $g$ and $h$ denote determinants of the metric and of the
static metric, and are not to be confused with the gravitational
acceleration and the rise of the bob in Section \ref{sec:close}.
A rope with negligible mass and extensibility has zero energy density,
$\rho=0$. Using $g=-N^{2}h$,
\begin{align}
0 & =\Gamma_{jk}^{i}S^{kj}+\frac{1}{N\sqrt{h}}\partial_{j}\left(N\sqrt{h}S^{ij}\right)\\
 & =\Gamma_{jk}^{i}S^{kj}+\frac{1}{\sqrt{h}}\partial_{j}\left(\sqrt{h}S^{ij}\right)+\frac{1}{N}S^{ij}\partial_{j}N\\
 & =\frac{1}{N}\nabla_{j}\left(NS^{ij}\right).
\end{align}
From now on, $\nabla$ is understood as the spatial Levi-Civita derivative;
$\Gamma_{jk}^{i}$ equals the Christoffel symbol of the spatial metric.
For $S^{ij}=-\ten\tv^{i}\tv^{j}$,
\begin{align}
0 & =\nabla_{j}\left(N\ten\tv^{i}\tv^{j}\right)\\
 & =\tv^{i}\nabla_{j}\left(N\ten\tv^{j}\right)+N\ten\tv^{j}\nabla_{j}\tv^{i}\label{eq:cons}
\end{align}
Contract with $\tv_{i}$ and use $\tv_{i}\tv^{i}=1$,
\begin{align}
0 & =\nabla_{j}\left(N\ten\tv^{j}\right)+N\ten\tv^{j}\tv_{i}\nabla_{j}\tv^{i}\text{ use metric compatibility}\\
 & =\nabla_{j}\left(N\ten\tv^{j}\right)+\frac{1}{2}N\ten\tv^{j}\nabla_{j}\left(\tv_{i}\tv^{i}\right)\\
 & =\nabla_{j}\left(N\ten\tv^{j}\right).\label{eq:divNT}
\end{align}
Plugging this back into \eqref{eq:cons}, and assuming non-vanishing
$N\ten$, gives
\begin{equation}
\tv^{j}\nabla_{j}\tv^{i}=0,
\end{equation}
thus proving that the rope follows a geodesic of the spatial metric.
Thus the massless taut rope is a geodesic of the spatial
metric~\eqref{eq:spatialmetric}.  The same spatial geodesics arise as the
Euler--Lagrange equations of the length functional
$\int \dd\ell$,
with \(\dd\ell\) defined by Eq.~\eqref{eq:spatialmetric}; this is the
standard  characterization of geodesics
\cite[Chap.~6, Thm.~6.3 and Cor.~6.5]{lee2019RiemManif}.

Although not needed below, integrating \eqref{eq:divNT} over a rope
segment shows that the redshifted tension $N\ten$, rather than $\ten$,
is constant along the rope.

\subsection{Angular displacement of the bob}
\label{sec:ang}
Initially the rope is straight and both ends lie at $\theta=0$. Its fixed proper length is
\begin{equation}
\LL=\int_{r_2}^{r_1 }\frac{\dd r}{N(r)}.
\label{eq:Lstraight}
\end{equation}
Now displace the bob slightly and hold it so that its $\theta$
coordinate 
is $\vartheta\ll1$. After transients die out, the rope becomes the
spatial geodesic joining $(r_1 ,0)$ to $(r_2+\delta r ,\vartheta)$, where
$\delta r>0$.
Because the metric does not depend on $\theta$, the 
geodesic has a conserved quantity
\begin{equation}
r^2\frac{\dd\theta}{\dd s}=J,
\label{eq:Jdef}
\end{equation}
where $J$ is constant along the rope (but changes when the position of
the bob changes). Combining Eq.~\eqref{eq:Jdef}
with the normalization of the arc length,
\begin{equation}
1=\frac{1}{N^2}\left(\frac{\dd r}{\dd s}\right)^2+r^2\left(\frac{\dd\theta}{\dd s}\right)^2,
\end{equation}
we obtain
\begin{equation}
\dd s=\frac{\dd r}{N\sqrt{1-\frac{J^2}{r^2}}}.
\label{eq:dsdr}
\end{equation}
The fixed proper length condition becomes
\begin{equation}
\LL=\int_{r_2+\delta r }^{r_1 }\frac{\dd r}{N\sqrt{1-\frac{J^2}{r^2}}}.
\label{eq:Ldeflected}
\end{equation}
The angular displacement of the bob is
\begin{equation}
\vartheta
= J\int_{0}^{L_0 }\frac{\dd s}{r^2}
=J\int_{r_2+\delta r }^{r_1 }\frac{\dd r}{r^2N\sqrt{1-\frac{J^2}{r^2}}}.
\label{eq:thetaexact}
\end{equation}
Since $\vartheta$ and $J$ are both small quantities of the same order,
we may neglect cubic and higher-order terms and write 
\begin{equation}
\vartheta\simeq J\int_{r_2}^{r_1 }\frac{\dd r}{r^2N}.
\label{eq:thetasmall}
\end{equation}
Using
\begin{equation}
\frac{\dd N}{\dd r}=\frac{r_s }{2r^2N},
\end{equation}
this integral is elementary and yields
\begin{equation}
\vartheta\simeq \frac{2J}{r_s }\left(N_1 -N_2\right).
\label{eq:thetaJ}
\end{equation}

\subsection{Radial lift of the bob}
\label{sec:lift}

Here we show that, when the pendulum is deflected by a small angle
$\vartheta$, the bob increases its radial coordinate by $\delta
r\sim\vartheta^2$. In the case of the usual classical pendulum of
length $\ell$ deflected by a small angle $\alpha$,   $\delta
r$ corresponds to the change of the elevation by
$\ell\left(1-\cos\alpha\right)\simeq\alpha^{2}\ell/2$.

To establish how $\delta r$ and $\vartheta$ are related, we compare Eqs.~\eqref{eq:Lstraight} and
\eqref{eq:Ldeflected}, which represent the same rope length. Expanding
to quadratic order in $J$ and linear order in $\delta r$ gives
\begin{align}
0
&=\int_{r_2+\delta r }^{r_1 }\frac{\dd r}{N\sqrt{1-\frac{J^2}{r^2}}}-\int_{r_2}^{r_1 }\frac{\dd r}{N}
\nonumber\\
&\simeq \int_{r_2}^{r_1 }\left(\frac{1}{N\sqrt{1-\frac{J^2}{r^2}}}-\frac{1}{N}\right)\dd r-\frac{\delta r}{N_2}
\nonumber\\
&\simeq \frac{J^2}{2}\int_{r_2}^{r_1 }\frac{\dd r}{r^2N}-\frac{\delta r}{N_2}.
\end{align}
The integral is the same as in Eq.~\eqref{eq:thetasmall}, so for the
endpoint angle $\vartheta$ used in this static construction,
\begin{equation}
\delta r=\frac{N_2J\vartheta}{2}
=\frac{N_2r_s }{4\left(N_1 -N_2\right)}\vartheta^2.
\end{equation}
Equivalently, in the dynamical problem below, for an instantaneous bob
coordinate $\theta(t)$ the same quadratic constraint is
\begin{equation}
\delta r(\theta)=\frac{N_2r_s }{4\left(N_1 -N_2\right)}\theta^2.
\label{eq:deltar}
\end{equation}
We shall use this relation to eliminate the $r$-dependence in the
Lagrangian in the next Section. Note that in the limit of a weak
field, $r_{1,2}\simeq r_2\gg r_s$, and short pendulum,
$\ell=r_1-r_2\ll r_{1,2}$, we reproduce the usual elevation of a
classical pendulum. Using $r_2\theta\simeq \ell\alpha$, $\delta
r=\theta^2r_2^2/\left(2\ell\right)=\alpha^2\ell/2$, as expected.

\section{Small-amplitude oscillations}
\label{sec:dynamics}
Here we derive the period of small oscillations, expressing it
 in terms of $r_1$ and $r_2$ in Section \ref{sec:N1N2}, and then in terms
 of $r_2$ and the rope length $L_0$ in Section \ref{subsec:period-length}.

\subsection{Fixed equilibrium positions of fulcrum and bob}\label{sec:N1N2}

Let $\theta(t)$ be the bob's angular displacement, and let $\vartheta$
denote its maximum value during a small oscillation. Thus
$\theta=\order{\vartheta}$, while $\vartheta$ itself is a fixed small
amplitude.  We use the Schwarzschild coordinate time $t$ as the
evolution parameter; dots denote derivatives with respect to $t$. 
The free point-particle Lagrangian for the bob, before imposing the rope constraint, is
\begin{align}
\mathcal{L}_0 \dd t
&=-mc^2\dd\tau
\nonumber\\
&=-mc\sqrt{N^2c^2\dd t^2-\frac{\dd r^2}{N^2}-r^2\dd\theta^2}.
\end{align}
Hence the coordinate-time Lagrangian, defined by $S_0=\int\mathcal{L}_0\,\dd t$, is
\begin{equation}
\mathcal{L}_0=-mc^2N\sqrt{1-\frac{\dot r^2}{N^4 c^2}-\frac{r^2\dot\theta^2}{N^2c^2}}.
\label{eq:LagrangianExact}
\end{equation}
Motion of the bob in  the $(r,\theta)$ plane is constrained by the
fixed-proper-length condition.  To
quadratic order in the angular displacement, it is the holonomic
constraint  derived in Eq.~\eqref{eq:deltar},
\begin{equation}
\chi(r,\theta)
\equiv
\frac{N_2r_s }{4\left(N_1 -N_2\right)}\theta^2-
\left(r-r_2\right)=0.
\label{eq:constraintTheta}
\end{equation}
This constraint may be imposed by introducing a
Lagrange multiplier $\Lambda(t)$,
\begin{equation}
\mathcal{L}_{\rm c}
=
\mathcal{L}_0+
\Lambda(t)\left[
\frac{N_2r_s }{4\left(N_1 -N_2\right)}\theta^2-
\left(r-r_2\right)
\right].
\label{eq:LagrangianConstrained}
\end{equation}
Variation with respect to $\Lambda$ enforces the rope constraint,
while variation with respect to $r$ determines the corresponding
constraint force. Since the constraint has already been solved to the
required order, we may equivalently substitute
\begin{equation}
r=r_2+\frac{N_2r_s}{4\left(N_1-N_2\right)}\theta^2
\label{eq:rConstraintTheta}
\end{equation}
into the free part of the Lagrangian and keep terms through quadratic order in the amplitude.

The order counting is
\begin{equation}
\theta=\order{\vartheta},\qquad
\dot\theta=\order{\vartheta},\qquad
r-r_2=\delta r=\order{\vartheta^2},\qquad
\dot r=\order{\vartheta^2}.
\label{eq:orderCountingTheta}
\end{equation}
Therefore the radial-velocity term is of fourth order in the oscillation amplitude and may be neglected in the harmonic approximation. Expanding Eq.~\eqref{eq:LagrangianExact} to quadratic order gives
\begin{equation}
\mathcal{L}_0=-mc^2N+\frac{mr^2}{2N}\dot\theta^2+
\order{\vartheta^4 }.
\label{eq:LagrangianExpanded1}
\end{equation}
Now write $r=r_2+\delta r$ and expand $N$ about $r_2$:
\begin{align}
N
&=\sqrt{1-\frac{r_s }{r_2+\delta r}}
\nonumber\\
&\simeq N_2+\frac{r_s  \delta r}{2r_2^2N_2}.
\label{eq:Nexpand}
\end{align}
Substituting Eq.~\eqref{eq:deltar} and retaining only quadratic terms in the amplitude yields the reduced constrained Lagrangian
\begin{align}
\mathcal{L}_{\rm red}
&=-mc^2N_2-mc^2\frac{r_s \delta r}{2r_2^2N_2}+\frac{mr_2^2}{2N_2}\dot\theta^2
\nonumber\\
&=-mc^2N_2-mc^2\frac{r_s ^2}{8r_2^2\left(N_1 -N_2\right)}\theta^2+\frac{mr_2^2}{2N_2}\dot\theta^2.
\label{eq:LagrangianFinal}
\end{align}
The first term in Eq.~\eqref{eq:LagrangianFinal} is constant and the remaining two terms have the form
of  the Lagrangian of a harmonic oscillator. Thus coordinate-time period is
\begin{equation}
T_t=4\pi\frac{r_2^2}{cr_s }\sqrt{\frac{N_1 -N_2}{N_2}}.
\label{eq:Tcoord}
\end{equation}
A static observer at the bob measures proper time $\dd\tau_{(2)}=N_2 \dd t$, and therefore measures the local period
\begin{equation}
T_{(2)}=N_2T_t=4\pi\frac{r_2^2}{cr_s }\sqrt{N_2\left(N_1 -N_2\right)}.
\label{eq:T2}
\end{equation}
This is our main result. A static observer at the
fulcrum measures a longer period,
\begin{equation}
T_{(1)}=\frac{N_1 }{N_2}T_{(2)}.
\label{eq:T1}
\end{equation}
The factor $N_1/N_2$
relates periods measured by the two static observers, whereas the
oscillator period itself depends on $N_2\left(N_1 -N_2\right)$.

\subsection{Period in terms of the proper length}
\label{subsec:period-length}

Equation~\eqref{eq:T2} is written in terms of the endpoint radii, or
equivalently in terms of the two lapse values $N_1$ and $N_2$.  Since
the invariant rope length $\LL$ is the prescribed constraint, it is
useful to eliminate $r_1$ in favor of $\LL$ and $r_2$.  From
\begin{equation}
N^2=1-\frac{r_s}{r}
\end{equation}
we have
\begin{equation}
r=\frac{r_s}{1-N^2},
\qquad
\frac{\dd r}{N}=\frac{2r_s\,\dd N}{(1-N^2)^2}.
\end{equation}
Therefore Eq.~\eqref{eq:Lstraight} becomes
\begin{equation}
\LL
=
r_s\left[F(N_1)-F(N_2)\right],
\label{eq:L0F}
\end{equation}
where
\begin{equation}
F(N)
=
\frac{N}{1-N^2}+\operatorname{arctanh}N .
\label{eq:Fdef}
\end{equation}
Derivative of this function is positive,
\begin{equation}
F^\prime(N)=\frac{2}{(1-N^2)^2}>0,
\end{equation}
so $F$ is increasing for $0<N<1$.  Thus, for fixed bob radius
$r_2$, or fixed $N_2=N(r_2)$, Eq.~\eqref{eq:L0F} determines $N_1$
uniquely,
\begin{equation}
N_1
=
F^{-1}\left(F(N_2)+\frac{\LL}{r_s}\right).
\label{eq:N1ofL0}
\end{equation}
$F^{-1}$ is also an increasing function.
Substitution into Eq.~\eqref{eq:T2} gives the period as an implicit
function of the proper rope length and the bob radius,
\begin{equation}
T_{(2)}(\LL,r_2)
=
4\pi\frac{r_2^2}{cr_s}
\sqrt{
N_2
\left[
F^{-1}\left(F(N_2)+\frac{\LL}{r_s}\right)-N_2
\right]
}.
\label{eq:T2L0}
\end{equation}
Since $F^{-1}$ is increasing, we see that, for fixed $r_2$, a longer rope
gives a longer period, like in non-relativistic mechanics.

  In the limit as $\LL\to\infty$,
one has $N_1\to1$, and therefore
\begin{equation}
T_{(2)}(\LL,r_2)
\longrightarrow
4\pi\frac{r_2^2}{cr_s}\sqrt{N_2(1-N_2)} ,
\label{eq:T2L0limit}
\end{equation}
in agreement with Eq.~\eqref{eq:T2} in the same limit.

\section{Limiting regimes}
\label{sec:limits}

\subsection{Newtonian limit}

In the weak-field regime where both ends of the pendulum are far
from the horizon,  $r_{1,2}\gg r_s $, we have
\begin{equation}
N_i \simeq 1-\frac{r_s }{2r_i }.
\end{equation}
Setting $r_2=R$ and $r_1 =R+\LL$, with $\LL\ll R$, Eq.~\eqref{eq:T2} becomes
\begin{align}
T_{(2)}
&\to 4\pi\frac{R^2}{cr_s }\sqrt{\frac{r_s }{2R}-\frac{r_s }{2(R+\LL)}}
\nonumber\\
&\simeq 2\pi\sqrt{\frac{R^2\LL}{GM}}
=2\pi\sqrt{\frac{\LL}{g}},
\end{align}
where $g=GM/R^2$. Thus the classical period is recovered.

\subsection{Close-to-horizon limit and effective length}
\label{sec:close}
Now consider the case of the bob being just above the horizon:
$r_2\to r_s $, so that  $N_2\ll1$, while the fulcrum stays at fixed
$r_1$ with $N_1\gg N_2$. Equation~\eqref{eq:T2} gives
\begin{equation}
T_{(2)}\simeq \frac{4\pi r_s }{c}\sqrt{N_1 N_2}.
\label{eq:T2NearHorizon}
\end{equation}
It is useful to reinterpret this result in terms of the geometry perceived
by a static observer at the bob. 
\begin{figure}[htb]
\centering
\includegraphics[width=0.5\textwidth]{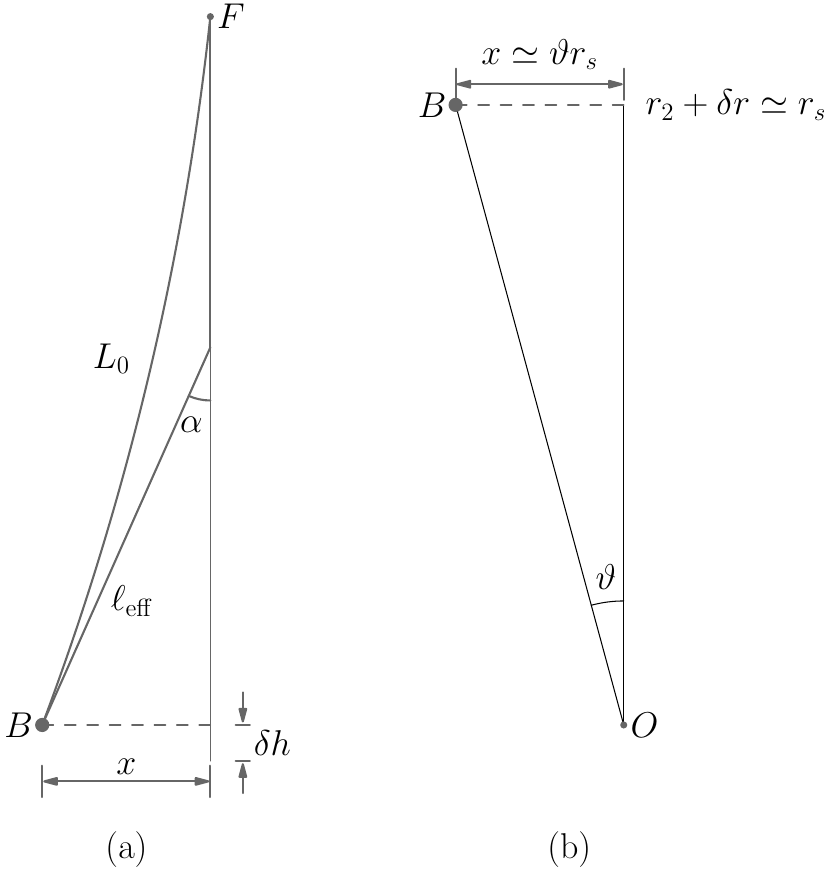}\par
\caption{Close-to-horizon geometry of the deflected
  pendulum with fulcrum $F$ and bob $B$. (a) A local static observer sees a horizontal displacement
  $x$, a vertical rise $\delta h$, and an effective length
  $\ell_{\mathrm{eff}}$. (b) $x$ can also be expressed in terms of
  $r_s$ and the polar deflection $\vartheta$. $O$ denotes the origin
  of the Schwarzschild coordinate system. $\alpha$ is the local
  effective-pendulum angle, analogous to the angle $\alpha$ mentioned in
  Sect.~\ref{sec:lift} in the context of a classical pendulum.}
\label{fig:nearhorizon}
\end{figure} %
Such observer sees a horizontal
displacement $x$ and vertical rise $\delta h$ (see
Fig.~\ref{fig:nearhorizon}; $\delta h$ is a physical distance and
differs from the coordinate difference $\delta r$ by a lapse factor,
see Eq.~\eqref{eq:deltah}). At a turning point, where
$|\theta|=\vartheta$, the horizontal displacement is 
\begin{equation}
x=r_2\vartheta\simeq r_s \vartheta,
\end{equation}
and
the corresponding rise of the bob is
\begin{equation}
\delta h=\frac{\delta r}{N_2}=\frac{r_s }{4\left(N_1 -N_2\right)}\vartheta^2
\simeq \frac{r_s }{4N_1 }\vartheta^2.
\label{eq:deltah}
\end{equation}
This interpretation assumes that $\delta r$ is
small compared with the coordinate distance from the horizon,
\begin{equation}
\delta r \ll r_2-r_s .
\end{equation}
Using Eq.~\eqref{eq:deltar}, this condition becomes
\begin{equation}
\vartheta^2 \ll
\frac{4N_2(N_1-N_2)}{1-N_2^2}
\simeq 4N_2(N_1-N_2),
\end{equation}
where the last form is the close-to-horizon limit. With this ordering
the expansion of \(N(r_2+\delta r)\) and the estimate
\(\delta h\simeq \delta r/N_2\) are uniform as we take $r_2 \to r_s$.

Now compare the behavior in Eq.~\eqref{eq:deltah} with  a Newtonian pendulum of 
length $\ell_{\mathrm{eff}}$. Its  small-angle rise is
\begin{equation}
\delta h=\frac{x^2}{2\ell_{\mathrm{eff}}}.
\end{equation}
Comparing with Eq.~\eqref{eq:deltah} gives
\begin{equation}
\ell_{\mathrm{eff}}=2N_1 r_s .
\label{eq:leff}
\end{equation}
Thus the local observer sees the pendulum behave as if its length were
not the full proper rope length $\LL$, but the shorter effective
length $\ell_{\mathrm{eff}}$. 

The local gravitational acceleration at the bob is
\begin{equation}
g=\frac{c^2r_s }{2r_2^2N_2}\simeq\frac{c^2}{2r_s N_2}.
\end{equation}
Therefore the Newtonian estimate based on the local quantities,
\begin{equation}
T_{\mathrm{class}}=2\pi\sqrt{\frac{\ell_{\mathrm{eff}}}{g}},
\end{equation}
becomes
\begin{equation}
T_{\mathrm{class}}=\frac{4\pi r_s }{c}\sqrt{N_1 N_2},
\label{eq:T2atrs}
\end{equation}
which agrees with Eq.~\eqref{eq:T2NearHorizon}. This local
interpretation is consistent with the fact that static observers very
close to the horizon are highly accelerated idealized probes rather than
inertial ones~\cite{Mazzitelli2020}. In particular, even for a very
long pendulum,  for which
$N_1$ tends to 1, the effective length, Eq.~\eqref{eq:leff}, is of the
order of the Schwarzschild radius. The very large gravitational
acceleration close to the horizon explains the smallness of the period
$T_{(2)}$ in Eq.~\eqref{eq:T2atrs}.

\section{Comparison with the coordinate-length model}
\label{sec:comparison}

\begin{figure}[htb]
  \centering
  \includegraphics[scale=.6]{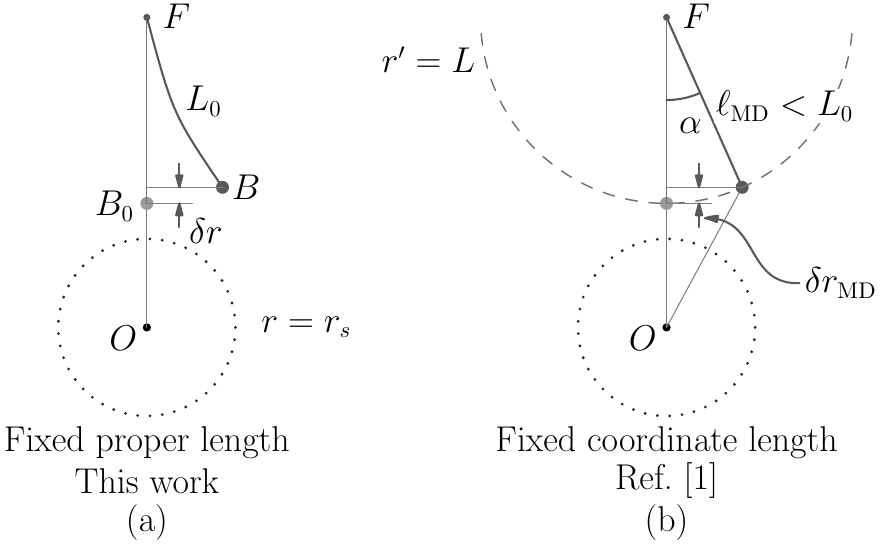}
  \caption{Comparison of the two pendulum constraints. The dotted circle around the
      origin $O$ denotes the event horizon. The gray dot denotes the
      equilibrium position $B_0$ of the bob $B$. (a) In the invariant model used
      in Sections \ref{sec:geometry} and \ref{sec:dynamics}, the
      displaced rope keeps its proper length $L_0$; the bob rises by
      $\delta r$ given in Eq.~\eqref{eq:deltar}. (b)
      In the coordinate-constraint model of
      Ref.~\cite{Martin-Delgado:2022nmx}, the bob's coordinate
      distance from the fulcrum $F$ is always $r'=L$ but its
      proper distance from $F$ shrinks to $\ell_{\rm MD}(\alpha)<L_0$,
      see Eq.~\eqref{eq:lmdnearhorizon}. Close to the
      horizon, the lift $\delta r_{\rm MD}$ in Eq.~\eqref{eq:MDlift} is dominated by this shortening.}
\label{fig:compare}   
\end{figure}

To compare  with the pioneering study of a Schwarzschild pendulum in Ref.~\cite{Martin-Delgado:2022nmx}, it is
convenient to adopt its notation. The fulcrum is placed at
Schwarzschild radial coordinate $\Rone $, and new coordinates
$(r^\prime ,\theta^\prime ,\phi^\prime )$ are introduced by
translating the
origin from the gravitating mass to the fulcrum:
\begin{equation}
r^2=\Rone ^2+r^{\prime 2}+2\Rone r^\prime \cos\theta^\prime ,
\qquad
\tan\theta=\frac{r^\prime \sin\theta^\prime }{\Rone +r^\prime \cos\theta^\prime },
\qquad
\phi=\phi^\prime .
\label{eq:MDcoords}
\end{equation}
The pendulum constraint in that work is then taken to be
\begin{equation}
r^\prime =L,
\end{equation}
namely a fixed coordinate distance from the fulcrum. Let
\begin{equation}
\rzero \equiv \Rone -L>r_s 
\end{equation}
be the bob's equilibrium Schwarzschild radial coordinate, and write the small
angular displacement from the downward vertical as (see Fig.~\ref{fig:compare}(b))
\begin{equation}
\theta^\prime =\pi-\alpha,
\qquad
\alpha\ll 1.
\end{equation}
Equation~\eqref{eq:MDcoords} gives
\begin{equation}
r^2=\rzero ^2+\Rone L \alpha^2+\order{\alpha^4 },
\qquad
r=\rzero +\delta r_{\rm MD}+\order{\alpha^4 },
\label{eq:MDlift1}
\end{equation}
with
\begin{equation}
\delta r_{\rm MD}=\frac{\Rone L}{2\rzero }\alpha^2.
\label{eq:MDlift}
\end{equation}
Thus a sideways displacement at fixed $r^\prime =L$ produces an
outward shift in the 
Schwarzschild radial coordinate, as expected.

The crucial point is that $r^\prime =L$ does not keep the rope length
fixed. The proper length of the coordinate segment joining the fulcrum
to the bob is
\begin{equation}
\ell_{\rm MD}(\alpha)=
\int_0^L
\dd r^\prime  
\sqrt{
\frac{1}{N(r)^2}
\left(\frac{\partial r}{\partial r^\prime }\right)^2
+r^2
\left(\frac{\partial \theta}{\partial r^\prime }\right)^2
},
\label{eq:lmdexact}
\end{equation}
where $r=r(r^\prime ,\alpha)$ and $\theta=\theta(r^\prime ,\alpha)$ follow from
Eq.~\eqref{eq:MDcoords}. Expanding to quadratic order gives
\begin{equation}
\ell_{\rm MD}(\alpha)
=
\LL
-\alpha^2\frac{r_s }{2}
\int_{\rzero }^{\Rone }\frac{\dd r}{N(r)}
\left[
\frac{\Rone ^2}{r^3 }
+\frac{\Rone (\Rone -r)}{2r^2( r - r_s )}
\right]
+\order{\alpha^4 }.
\label{eq:lmdexpand}
\end{equation}
The integral is strictly positive, so the proper length of the
coordinate segment decreases as soon as the bob is moved sideways.
Close to the horizon, the lower end of the integral dominates and
\begin{equation}
\ell_{\rm MD}(\alpha)-\LL
=-\frac{\Rone L}{2\sqrt{r_s (\rzero -r_s )}}\alpha^2
+\order{\alpha^2},
\qquad
\rzero \to r_s ,
\label{eq:lmdnearhorizon}
\end{equation}
where the omitted terms are less singular in $\rzero -r_s $. Since
$\Nrzero\simeq\sqrt{(\rzero -r_s )/r_s }$ in the same limit,
Eq.~\eqref{eq:MDlift} implies
\begin{equation}
\frac{\delta r_{\rm MD}}{\Nrzero}
=\frac{\Rone L}{2\sqrt{r_s (\rzero -r_s )}}\alpha^2
+\order{\alpha^2}.
\label{eq:lmdshift}
\end{equation}
Thus, in the strong-field regime, the radial lift is dominated by
compensation for the shortening of the rope's proper length.
Rather than an inextensible pendulum in the usual
sense, it is a different constrained system, closer to a bead forced
to remain on the coordinate sphere $r^\prime =L$.

With $r^\prime =L$ understood as a holonomic constraint, the small-oscillation
frequency is obtained from the reduced Lagrangian. Using
\begin{equation}
r=\rzero +\frac{\Rone L}{2\rzero }\alpha^2+\order{\alpha^4 },
\qquad
\theta=\frac{L}{\rzero }\alpha+\order{\alpha^3 },
\end{equation}
one finds
\begin{equation}
\mathcal{L}_{\rm MD}
=-mc^2\Nzero 
-mc^2\frac{r_s \Rone L}{4\rzero ^3 \Nzero }\alpha^2
+\frac{mL^2}{2\Nzero }\dot\alpha^2
+\order{\alpha^4 }.
\label{eq:Lmdred}
\end{equation}
The coordinate-time frequency is
therefore
\begin{equation}
\omega_{\rm MD}^2
=c^2\frac{r_s \Rone }{2L\rzero ^3 },
\qquad
T_{{\rm MD},(2)}
=\Nzero \frac{2\pi}{\omega_{\rm MD}}
=\frac{2\pi}{c}
\sqrt{\frac{2L\rzero ^3 \Nzero ^2}{r_s \Rone }}.
\label{eq:Tmd}
\end{equation}
In the weak-field limit $r_s \ll \rzero $ and $L\ll \Rone $,
Eq.~\eqref{eq:Tmd} agrees with our Eq.~\eqref{eq:T2}; indeed,
$\NRone -\Nzero \simeq r_s L/(2\Rone \rzero )$. Away from that limit the two
models are inequivalent, and in particular they predict different
close-to-horizon scalings,
\begin{equation}
T_{{\rm MD},(2)}\propto \Nzero ,
\qquad
T_{(2)}\propto \sqrt{\Nzero }.
\end{equation}

Two brief technical remarks are worth recording. First,
Appendix~B of Ref.~\cite{Martin-Delgado:2022nmx} omits a factor of
$g^\prime _{12}$ in $\Gamma^{\prime 2}{}_{22}$; the correct expression is
\begin{equation}
\Gamma^{\prime 2}{}_{22}
=
\frac{-g^\prime _{12}\left(2\partial_{\theta^\prime }g^\prime _{12}-\partial_{r^\prime }g^\prime _{22}\right)
+g^\prime _{11}\partial_{\theta^\prime }g^\prime _{22}}
{2\left[ g^\prime _{11}g^\prime _{22}-(g^\prime _{12})^2\right]}.
\label{eq:Gamma222correct}
\end{equation}
Since $g^\prime _{12}=0$ at equilibrium but
$\partial_{\theta^\prime }g^\prime _{12}\neq0$, omitting that factor changes the
linearized equation. Second, the weak-field estimate
$E/c^2\simeq1-r_s /(2|\Rone -L|)$ used in
Ref.~\cite{Martin-Delgado:2022nmx} is then carried into regimes with
$|\Rone -L|\sim r_s $, beyond its range of validity.

\section{Consistency of the fixed length assumption}
\label{sec:rigidity}

The fixed proper length of the rope is an idealization, plausible as a quasi-static
approximation. Changes of stress propagate slower than light and cannot affect the whole rope immediately.

In addition to the fixed proper length, we also assume that the rope adjusts
quasi-statically as the bob moves: it always follows a geodesic of the
spatial metric. This is valid only if
a signal can travel along the rope in much shorter time than the oscillation period.
The resulting condition on the rope length is derived in the remainder of this section.
If it is not satisfied, the fixed-length model must be replaced by
a model of an elastic string with finite
signal-propagation speed, accounting for  internal modes and phase
delays. Such an analysis is beyond our present scope. Difficulties
with the notion of relativistic rigid bodies are discussed in
Ref.~\cite[Sect.~15]{Landau:1975pou}.  (An
inextensible string is pedagogically discussed in
Ref.~\cite{Hankin2021} in a different, static context.)

Let
\begin{equation}
r_2=r_s(1+\epsilon),
\qquad
r_1=r_s(1+\xi),
\qquad
0<\epsilon<\xi ,
\end{equation}
and consider the bob close to the horizon, so that $\epsilon \ll 1$ and
\begin{equation}
N_2=\sqrt{1-\frac{1}{1+\epsilon}}\simeq \sqrt{\epsilon} .
\end{equation}
A radial light signal sent
from the fulcrum to the bob and back has round-trip proper time,
measured at the fulcrum (this is a special case of the Shapiro time
delay calculation, see for example \cite[Sect.~3.6.5]{GourgoulhonGR}),
\begin{equation}
\tau_{\rm sig}^{(1)}
= \frac{2N_1}{c}
\int_{r_2}^{r_1}\frac{\dd r}{N(r)^2}
=
\frac{2r_s}{c}N_1
\left(
\xi-\epsilon+\ln\frac{\xi}{\epsilon}
\right).
\end{equation}
For a real rope the signal may be a tension wave, slower than
light. We take $c$ as the propagation speed   since our goal is only
to explore the range of lengths in which a pendulum is in principle
conceivable.

The period measured at the fulcrum is given in
Eq.~\eqref{eq:T1}, $T_{(1)} = 4\pi r_2^2 N_1 \sqrt{N_1/N_2-1}/(c r_s)$.
The fixed-length approximation is consistent only if the signal can
travel many times during one period, 
\begin{equation}
\frac{\tau_{\rm sig}^{(1)}}{T_{(1)}}
\simeq
\frac{
\xi-\epsilon+\ln(\xi/\epsilon)}
{
2\pi \sqrt{N_1/N_2-1}
}
\ll 1 .
\label{eq:causal-ratio-general}
\end{equation}
For fixed finite
fulcrum position \(r_1\), the signal time grows only logarithmically as
\(\epsilon\to 0\), while the period measured at the fulcrum grows as a power,
$ T_{(1)}\propto \epsilon^{-1/4}$.
Therefore, for a fixed fulcrum position, our fixed-length model
improves as the bob approaches the horizon,
\begin{equation}
\frac{\tau_{\rm sig}^{(1)}}{T_{(1)}}\to0
\quad
\text{as}
\quad
\epsilon\to0 .
\end{equation}
The reason is that, from the fulcrum's point of view, the bob's motion becomes
increasingly redshifted.

For a very distant fulcrum, \(\xi\gg1\), we have \(N_1\simeq1\), and
Eq.~\eqref{eq:causal-ratio-general} gives
\begin{equation}
\frac{\tau_{\rm sig}^{(1)}}{T_{(1)}}
\simeq
\frac{\epsilon^{1/4}}{2\pi}
\left(
\xi+\ln\frac{\xi}{\epsilon}\right).
\end{equation}
The boundary where the signal time is comparable to the period is therefore
\begin{equation}
\xi+\ln\frac{\xi}{\epsilon}
\sim
2\pi \epsilon^{-1/4}.
\end{equation}
To leading order, the logarithm is smaller than the power-law term, giving
\begin{equation}
\xi_{\max}\sim 2\pi \epsilon^{-1/4}.
\end{equation}
Equivalently, the rough consistency condition is
\begin{equation}
r_1-r_s
\ll
2\pi r_s\epsilon^{-1/4}.
\end{equation}
The important point is
that the allowed distance of the fulcrum grows as \(\epsilon\) 
decreases, because the period measured at the fulcrum grows faster than the
signal delay.

There is also a  close-to-the-horizon subcase in which the fulcrum approaches
the horizon together with the bob. Let
\begin{equation}
\xi=\lambda\epsilon,
\qquad
\lambda>1,
\end{equation}
with \(\lambda\) fixed while \(\epsilon\to0\). Then
\begin{equation}
N_2\simeq\sqrt{\epsilon},
\qquad
N_1\simeq\sqrt{\lambda \epsilon}.
\end{equation}
Substituting into Eq.~\eqref{eq:causal-ratio-general} gives
\begin{equation}
\frac{\tau_{\rm sig}^{(1)}}{T_{(1)}}
\to
\frac{\ln\lambda}
{2\pi\sqrt{\sqrt{\lambda}-1}} .
\end{equation}
The right-hand side vanishes at the boundaries of its range, $1\leq \lambda
< \infty$, and has  a maximum
\begin{equation}
\max_{\lambda>1}
\left[
\frac{\ln\lambda}
{2\pi\sqrt{\sqrt{\lambda}-1}}
\right]
\simeq 0.256
\text{ for } \lambda\simeq 24.2 .
\end{equation}
Thus, when both the bob and fulcrum approach the horizon with fixed
\(\xi/\epsilon=\lambda\), the light-crossing time stays finite relative to the
period and never becomes larger than it. However,  it is {\em much}
smaller than the period  only for $\lambda$
close to 1 as well as for very large $\lambda$.

\section{Conclusions}
\label{sec:conclusions}

We have determined the small-oscillation period of a pendulum in Schwarzschild spacetime under an
invariant constraint: the rope is taut, very light, inextensible, and of fixed
proper length. The displaced rope follows a geodesic of the spatial
metric on a static Schwarzschild slice. 
These assumptions  lead to the period measured by a static observer at the bob,
\begin{equation}
T_{(2)}=4\pi\frac{r_2^2}{cr_s}\sqrt{N_2\left(N_1-N_2\right)}.
\end{equation}
The result passes two consistency checks. In the weak-field, short-rope limit it
reduces to the Newtonian period \(2\pi\sqrt{\LL/g}\). In the close-to-horizon
regime it has a simple interpretation: the pendulum behaves as if it had
an effective length \(\ell_{\mathrm{eff}}=2N_1 r_s\), and when this
reduced length is combined
with the large gravitational acceleration perceived by a static observer close to the horizon,
one recovers the same limiting period.

The causal analysis in Section \ref{sec:rigidity} determines limits of the fixed-length
approximation. The rope may be treated as adjusting quasi-statically only when
the signals can travel  along it in  much shorter time than the oscillation period.
Close to the horizon this condition is less restrictive than far from
it because the bob's motion  is strongly
redshifted, but for a long rope or a material with small speed of
tension waves, the dynamics of the rope cannot be ignored.

The main difference with Ref.~\cite{Martin-Delgado:2022nmx} is
geometric. We impose a fixed proper-length constraint, whereas that
work fixes a coordinate radius. Away from the weak-field limit the
latter prescription requires the proper length of the rope to vary,
and close to the horizon the apparent lift is dominated by that
shortening. The coordinate-length model yields Eq.~\eqref{eq:Tmd},
which agrees with our result only in the weak-field regime and
predicts a different close-to-horizon scaling.

The pendulum provides an example of how spatial geometry and redshift
enter mechanics in curved spacetime. Extensions of the present work
could relax the small-amplitude approximation or include the rope's
own energy density and backreaction, which may become important close
to the horizon.  One could also include effects of the gravitational
radiation emitted by the pendulum, together with the resulting damping
of the oscillations.  An extension to rotating black holes would
require a different formulation, beyond our static-slice construction.

\section*{Acknowledgements}

This research was supported by Natural Sciences and Engineering
Research Council of Canada (NSERC). We used Asymptote \cite{Asymptote}
to draw the figures.

\end{document}